# Use of Eye-Tracking Technology to Investigate Cognitive Load Theory

Tianlong Zu,[1] John Hutson,[2] Lester C. Loschky,[2] and N. Sanjay Rebello[1,3]

[1]*Department of Physics and Astronomy, Purdue University, 525 Northwestern Ave., West Lafayette, IN 47907*
[2]*Department of Psychological Sciences, Kansas State University, 492 Bluemont Hall, Manhattan, KS 66506*
[3]*Department of Curriculum and Instruction, Purdue University, 100 N. University St., West Lafayette, IN 47907*

Cognitive load theory (CLT) provides us guiding principles in the design of learning materials. CLT differentiates three different kinds of cognitive load -- intrinsic, extraneous and germane load. Intrinsic load is related to the learning goal, extraneous load costs cognitive resources but does not contribute to learning. Germane load can foster learning. Objective methods, such as eye movement measures and EEG have been used measure the total cognitive load. Very few research studies, if any, have been completed to measure the three kinds of load separately with physiological methods in a continuous manner. In this current study, we will show how several eye-tracking based parameters are related to the three kinds of load by having explicit manipulation of the three loads independently. Participants having low prior knowledge regarding the learning material participated in the study. Working memory capacity was also measured by an operation memory span task.

## I. INTRODUCTION & THEORY

Cognitive load theory (CLT) provides us guiding principles for designing instructional material [1, 2]. Human beings have a long-term memory which can store extremely large amount of information and a working memory which has finite capacity [3]. Processing of the information happens in working memory, knowledge is said to be gained only after it is stored in long-term memory. Optimal instructional design should impose cognitive load that does not exceed the learner's working memory capacity.

CLT differentiates three different kinds of load which are assumed to be additive. Intrinsic cognitive load (ICL) refers to the mental effort needed to learn a concept. It is not affected by instructional design rather is related to the difficulty of the material. Extraneous cognitive load (ECL) depends upon the presentation format (e.g. visual, audio, text) as well as spatial and temporal organization. For example, if the same information is presented simultaneously in both text and audio, the redundancy costs cognitive resources which increases the. extraneous load, and hinders learning. Germane cognitive load (GCL) relies on students' motivation. Since motivation affects level of engagement of students during learning. GCL is referred as the mental resources devoted to processing relevant information from learning material. It helps build and automate schemas to facilitate learning. Optimal design entails low ECL and high GCL

As online learning becomes ubiquitous, continuously monitoring each type of cognitive load, using real-time, non-intrusive, objective methods is important. Existing methods do not satisfy these requirements [4]. Eye-tracking methods show promise because they can potentially continuously and non-intrusively monitor students in real time. They have also been shown to differentiate tasks imposing low and high cognitive loads [5]. However, to date they have not been used to differentiate between different types of load. Our research question is: What eye-movement parameters are sensitive to different types of cognitive load?

## II. DESIGN OF STUDY

We manipulated the three types of cognitive load following a design by DeLeeuw and Mayer [6]. N=51 students (48 females, 3 males) enrolled in a physics class for elementary teachers were paid $30 each to participate in this 90-minute long study. The learning material was a multimedia lesson explaining how electric motor works. Figure 1 shows a screen shot of the learning materials.

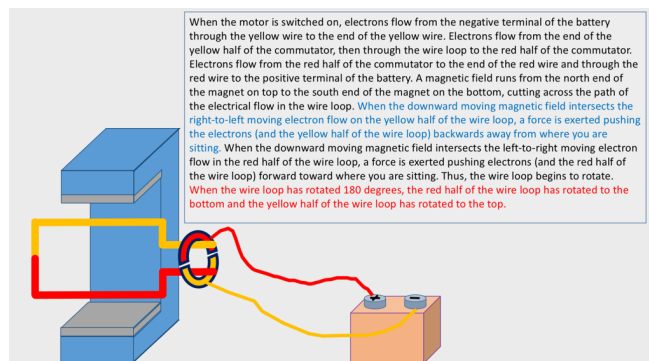

FIG 1. Screen shot of the multimedia lesson

ICL was manipulated in terms of sentence complexity which determines how many concepts need to be processed simultaneously. This was a within-subject factor, with the multimedia lesson consisting of four high-complexity and four low-complexity sentences. Example of high/low-complexity sentences can be found in figure 1 as highlighted in blue and red.

ECL was manipulated by redundancy effect [7]. Participants were randomly assigned to one of two conditions. In the non-redundant condition, they were presented a lesson version with animation and narration. In the redundant condition, students were presented a lesson version with animation, narration, and concurrent onscreen text that replicated the narration. This redundancy increases ECL.

GCL depends on learners' interaction with the learning material, and consequently their performance on a transfer test was used to designate them as having high or low GCL. Transfer tasks included seven conceptual questions presented in a fixed order. Students had three minutes to answer each of the seven questions. Students whose scores were higher than the mean on the transfer task were categorized as having high GCL. The rest were categorized as having low GCL. This is hypothesized as better scoring performance directly benefits from high GCL.

The study procedure is shown in Fig. 2. A short pre-survey was given to ascertain participants' prior knowledge about the material. They rated their knowledge level of each question on a Likert scale from 1 (I don't know anything about this image) to 5 (I understand this image very well). Written answers to each question were also collected. Individual difference in working memory capacity (WMC) was identified as a potential confounding variable in cognitive load study. There are several existing methods to measure human working memory capacity. We chose an easy-to-conduct automatic operation span task to measure WMC [8]. This operation span task consists of 15 sets of tasks. Each set contains 3 to 7 arithmetic operations. A letter at the end of each arithmetic operation is to be memorized by participants. At the end of each set, students need to report back the memorized letters in the order they were presented by selecting from a letter matrix on the computer screen. Data will be used only if the operation processing component reaches 85% correctness rate. Working memory capacity is reflected by the total number of letters reported correctly.

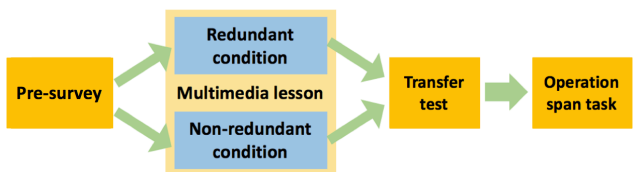

FIG 2. Interview Procedure

Measurements of cognitive load included three measures used by DeLeeuw and Mayer: (i) Response time to a secondary task which asked participants to press the spacebar immediately after noticing the screen background color starting to change from light gray to purple. (ii) Participants' subjective mental effort rating on a scale from 1 (extremely low) to 9 (extremely high) at the end of each selected sentence after spacebar is pressed. (iii) Participants' subjective difficulty level rating of the learning material on a scale from 1 (extremely easy) to 9 (extremely difficult) at the end of entire lesson. In addition to these measures, we also recorded participants' eye-movements during the learning phase using an EyeLink 1000 + eye-tracker.

## III. RESULTS

### A. Pre-Survey, Transfer Test and O-span Task

The mean (± S.D.) *pre-survey* students rating of their understanding of the learning material was 1.29±0.67. which indicates that they had low prior knowledge about electric motors. After examination, their written answers confirmed most of the students had minimal understanding toward the mechanism of electric motors.

The mean (± S.D.) *transfer test* score was 7.14±3.20. A mean-split rendered 28 students (below the mean) in low-germane load group and 18 students (above the mean) in high-germane load group.

The O-span task scores were normally distributed with a mean (± S.D.) of 32.72±15.90, with half students scoring above the mean and half below the mean.

### B. Non-Eye-Tracking Based Measures

*Response Time to Secondary Task*: A 2 x 2 mixed ANOVA was conducted to test the effect of within-subject factor (sentence complexity) and between-subject factor (redundancy) on response time to a secondary task. We found that there is a significant main effect of redundancy on response time to a secondary task, $F(1, 44) = 4.985$, $p = .031$, $\eta_p^2 = .102$. Students from redundant condition (high ECL) took longer time to press spacebar than those from non-redundant (low ECL) condition. No significant main effect of sentence complexity was found. An independent samples t-test conducted to compare the response time of students from low- and high-GCL groups found no significance, $t(44) = -0.400$, *ns*. (Means are presented in Table 1).

*Mental Effort Rating*: A 2 x 2 mixed ANOVA was conducted to test the effect of within-subject factor (sentence complexity) and between-subject factor (redundancy) on subjective mental effort rating. We found that there is a significant main effect of redundancy on mental effort rating, $F(1, 44) = 6.652$, $p = .013$, $\eta_p^2 = .102$. Students from redundant condition reported higher mental effort than those from non-redundant condition. We also found a significant main effect of sentence complexity, F(1, 44) = 24.9, $p < .001$, $\eta_p^2 = .361$., with participants rating their mental effort as higher when high-complexity sentences were processed. An independent samples t-test conducted to compare the means of subjective mental effort rating by students from low- GCL and high-GCL groups found no significance, $t(44) = .223$, ns. (Means are presented in Table 1).

*Difficulty Rating*: We collected this data only at the end of the learning phase. An independent samples t-test to test the effect of redundancy and transfer performance found no significant differences between difficulty ratings by students from redundant and non-redundant conditions, $t(44) = .423$, ns. A significant difference in means of difficulty rating between students from low- and high-GCL load groups was found, $t(44) = 2.157$, $p = .037$. Students from high-GCL group rated the material as less difficult than those from low-GCL group. (Means are presented in Table 1).

Our results as shown in Table 1 are consistent with results from DeLeeuw and Mayer [6].

TABLE 1. Means and SDs for three types of cognitive load manipulations based on three measures of cognitive load.

| Measures of Cognitive Load | Extraneous Load: Redundancy | |
|---|---|---|
| | Redundant | Non-redundant |
| Response Time | 1245(±512) | 968(±305) |
| Effort Rating | 5.45(±1.67) | 4.41(±1.51) |
| Difficulty Rating | 5.39(±1.50) | 5.22(±1.28) |
| | Intrinsic Load: Complexity | |
| | High | Low |
| Response Time | 1113(±512) | 1101(±446) |
| Effort Rating | 5.17(±1.47) | 4.70(±1.48) |
| | Germane Load: Transfer | |
| | High (n = 18) | Low (n = 28) |
| Response Time | 1139(±515) | 1086(±393) |
| Effort Rating | 4.88(±1.71) | 4.97(±1.27) |
| Difficulty Rating | 4.78(±1.59) | 5.64(±1.13) |

### C. Eye-Tracking Based Measures

To address the research question, several eye movement based parameters were examined against different types of cognitive load. Here we report a subset of them.

*Percentage Dwell Time*: In the learning material, there are different areas of interests (AoI): Animation and text AoIs in redundant condition and animation AoI in non-redundant condition. The percent of total time spent in a given AoI is a measure of the visual attention to that AoI.

First, we investigate whether adding text to animation and narration would shift students' attention. An independent samples t-test ($t(366) = 29.021$, $p < .001$.) found that the percentage dwell time spent looking at animation by students from non-redundant i.e. low ECL condition (86%) was significantly higher than that of those from redundant i.e. high ECL condition (38%). This means adding text will shift students' attention from animation to the text. Further, in the redundant condition, we found that students from high-GCL group had a significantly higher ($t(182) = -5.752$, $p < .001$) percentage dwell time (47%) in the animation AoI than those from low-GCL group (31%). This result is consistent with the redundancy effect. [7]. To summarize, percentage dwell time spent looking at animation AoI is sensitive to ECL as well as GCL.

*Mean Fixation Duration (MFD)*: Fixation duration records how long students look at one location where meaningful information exists. A hierarchical regression analysis was conducted to examine correlation between independent variables (IV) and the dependent variable *MFD*. In the first model, IVs include ICL, ECL and GCL. ICL, ECL and GCL are all categorical variables labeled by 0 and 1 representing low and high levels. We add o-span score as an IV to the second model. We found that ECL can uniquely predict *MFD* in the first model. It is worth noting that GCL also can uniquely predict MFD in addition to ECL and O-span in the second model. Results are in Table 2.

TABLE 2. Hierarchical regression for *MFD*

| IV | Model 1 | | Model 2 | |
|---|---|---|---|---|
| | β | p | β | p |
| ICL | .049 | .268 | .049 | .262 |
| ECL | -.547*** | .000 | -.549*** | .000 |
| GCL | .054 | .219 | .099* | .031 |
| O-SPAN | | | -.143** | .002 |

N = 46. $R^2 = .30^{***}$ for Model 1; $\Delta R^2 = .02^{**}$ for Model 2; * p < .05. ** p < .01. *** p < .001

*Mean Saccade Peak Velocity (MSPV)*: When students shift their attention from one fixating location to the next one, they make a saccade with non-uniform velocity. We conducted a hierarchical regression analysis for *mean saccade peak velocity*, using the same two models as before. We found that GCL and o-span score can predict *MSPV*. Results are in Table 3.

TABLE 3. Hierarchical Analysis for *MSPV*

| IV | Model 1 | | Model 2 | |
|---|---|---|---|---|
| | β | p | β | p |
| ICL | -.035 | .496 | -.035 | .49 |
| ECL | .096 | .06 | .093 | .065 |
| GCL | .214*** | .000 | .269*** | .000 |
| O-SPAN | | | -.176** | .001 |

N = 46. $R^2 = .06^{***}$ for Model 1; $\Delta R^2 = .03^{**}$ for Model 2; ** p < .01. *** p < .001.

*Ratio of Pupil Size Change (RPSC)*: This is defined as $(PS - PS_{baseline})/PS_{baseline}$ where $PS_{baseline}$ is the pupil size when no information is presented to students. Change of pupil size during learning can be said

TABLE 4. Hierarchical analysis for *RPSC*

| IV | Model 1 β | Model 1 p | Model 2 β | Model 2 p |
|---|---|---|---|---|
| ICL | .005 | .929 | .005 | .929 |
| ECL | .13* | .013 | .129* | .013 |
| GCL | -.131* | .012 | -.118* | .032 |
| O-SPAN | | | -.042 | .442 |

Note. N = 46. $R^2 = .03^{**}$ for Model 1; $\Delta R^2$ is ns for Model 2; * p < .05. ** p < .01

to be affected by cognitive load. We conducted a hierarchical regression analysis for *RPSC*, using the same two models as before. We found that GCL, ECL and o-span score all can uniquely predict *RPSC*. Results are in Table 4.

*Transition between Text & Animation AoIs (TBTA)*: This parameter records how many saccades were made between the two AoIs. This analysis is only done for the redundant condition, so it cannot be completed with respect to ECL. We conducted a hierarchical regression analysis for *TBTA*, using the same two models as before, except without ECL as an IV. We found that only ICL can uniquely predict *TBTA*. Results are in Table 5.

TABLE 5. Hierarchical regression for *TBTA*

| IV | Model 1 β | Model 1 p | Model 2 β | Model 2 p |
|---|---|---|---|---|
| ICL | .351** | .000 | .351*** | .000 |
| GCL | -.061 | .381 | -.087 | .226 |
| O-SPAN | | | -.099 | .170 |

Note. N = 46. $R^2 = .127^{***}$ for Model 1; $\Delta R^2 = .009$ is ns for Model 2; *** p < .001

## IV. CONCLUSIONS

Our results showed findings consistent with DeLeeuw & Mayer [6] in terms non-eyetracking based measurements, namely, a *subjective mental effort rating* was most sensitive to intrinsic load; *response time to a secondary task* was most sensitive to extraneous load; *subjective difficulty rating* was most sensitive to germane load.

To address the sensitivity of several eye movement based parameters to different types of cognitive load, our results showed that *mean fixation duration* was most sensitive to extraneous load (reflected by longer mean fixation duration for students from non-redundant condition than those from redundant condition); *mean saccade peak velocity* was most sensitive to germane load (reflected by higher value of peak velocity for high-transfer students than for low-transfer students); *transition between text & animation AoI* was most sensitive to intrinsic load (reflected by more transitions between text and animation AoIs made by students for high-complexity videos than for low-complexity videos); *Mean ratio of pupil size change* was most sensitive to extraneous and germane load (reflected by higher ratio of pupil size change of students from redundant and low-transfer condition/group than that of those from non-redundant and high-transfer condition/group).

We also found that in the redundant (high ECL) condition, adding concurrent onscreen text shifts students' attention from animation to text. Probably processing text is more automatic than processing animation when it comes to learning. Further we also found that high-GCL participants i.e. those who performed above the mean on the transfer test, had a higher dwell time in the animation region than in the text region. It appears that these students did not need to attend to the text to learn the concepts, rather they could attend to the animation and listen to the narration to learn from the multimedia lesson.

In educational settings, these findings can be used to monitor students' levels of the three kinds of cognitive load. Especially, use of eye-tracking technology allows instructors/educators to monitor that in real-time and continuous manner. It brings forward the possibility of integrating realtime feedback, adaptability and personalization of learning process.

## ACKNOWLEDGEMENTS

This work is supported in part by the U.S. National Science Foundation grant 1348857. Opinions expressed are of the authors and not necessarily of the Foundation. We grateful to Dr. R. E. Mayer and Dr. J. Sweller for insightful discussions and feedback to improve our study design.